\documentclass{iau} 

\usepackage{graphicx}	
\usepackage{amsmath}	
\usepackage{amssymb}	
\usepackage{upgreek}
\usepackage{booktabs}
\usepackage{mathrsfs}
\usepackage{calrsfs}
\usepackage{hyperref}
\hypersetup{colorlinks=false}
\usepackage{pdflscape}
\usepackage{rotating}
\usepackage{longtable}
\usepackage{multicol}

\title[Studying Magnetic Fields using Low-frequency Pulsar Observations] 
{Studying Magnetic Fields using Low-frequency Pulsar Observations}

\author[C. Sobey et al.]   
{C. Sobey$^{1,2}$
 \and LOFAR and MWA collaborations}

\affiliation{$^1$International Centre for Radio Astronomy Research - Curtin University, GPO Box U1987, Perth, WA 6845, Australia\\[\affilskip]
$^2$ CSIRO Astronomy and Space Science, 26 Dick Perry Avenue, Kensington, WA 6151, Australia\\email: {\tt c.sobey@curtin.edu.au}}

\pubyear{2017}
\volume{337}  
\jname{Pulsar Astrophysics -­ The Next 50 Years}
\editors{P. Weltevrede, B.B.P. Perera, L. Levin Preston \& S. Sanidas, eds.}

\begin{document}

\maketitle

\begin{abstract}

Low-frequency polarisation observations of pulsars, facilitated by next-generation radio telescopes, provide powerful probes of astrophysical plasmas that span many orders of magnitude in magnetic field strength and scale: from pulsar magnetospheres to intervening magneto-ionic plasmas including the ISM and the ionosphere. Pulsar magnetospheres with teragauss field strengths can be explored through their numerous emission phenomena across multiple frequencies, the mechanism behind which remains elusive. Precise dispersion and Faraday rotation measurements towards a large number of pulsars probe the three-dimensional large-scale (and eventually small-scale) structure of the Galactic magnetic field, which plays a role in many astrophysical processes, but is not yet well understood, especially towards the Galactic halo. We describe some results and ongoing work from the Low Frequency Array (LOFAR) and the Murchison Widefield Array (MWA) radio telescopes in these areas. These and other pathfinder and precursor telescopes have reinvigorated low-frequency science and build towards the Square Kilometre Array (SKA), which will make significant advancements in studies of astrophysical magnetic fields in the next 50 years.

\end{abstract}

\firstsection 
\section{Introduction}

Magnetic fields are ubiquitous throughout the Universe and play a role in numerous astrophysical processes across a range of physical scales. 
Pulsar magnetospheres with $\sim$teragauss field strengths contribute to generating electromagnetic radiation across the spectrum.  
The geomagnetic field with $\sim$1\,G field strengths protects the Earth from the solar wind that would otherwise erode the atmosphere.  
Galactic magnetic fields pervade the diffuse ISM with $\sim\upmu$G field strengths and influence many processes in the Galactic ecosystem, e.g., molecular-cloud and star formation and cosmic ray acceleration. 

We study magnetic fields in astrophysical plasmas through observing their effects on radiation. For example, the magnetic field direction and strength in sunspots was identified through Zeeman splitting in Solar spectra (Hale 1908). The Galactic magnetic field (GMF) was first measured using polarisation of starlight (Hall 1949; Hiltner 1949). Despite decades of study, our understanding of these magnetic fields is still relatively inadequate: the mechanism by which the pulsar magnetosphere generates and varies its emission remains obscure; and measurements of the 3-D structure of the GMF are sparse, especially towards the halo. However, work towards better understanding magnetic fields is ongoing and has been reinvigorated by the recent construction of next-generation low-frequency radio telescopes, including LOFAR and the MWA.

Here, we focus on observing the (polarised) emission from pulsars to study the pulsar magnetosphere (Section 2) and the Galactic magnetic field (Section 3).  We summarise and look forward to the next 50 years in Section 4.

\section{Studying the pulsar magnetosphere}

Magnetospheric emission from pulsars has been detected across the entire magnetic spectrum: from low radio frequencies ($\sim$10\,MHz; e.g. Hassall et al. 2012) up to high-energy gamma rays ($>$100\,GeV; e.g. Aliu et al. 2011). 
Despite decades of observations, the exact radio emission mechanism remains elusive. Low-frequency polarisation observations provide valuable insights into this, in combination with higher frequency data, e.g., spectral turnovers (e.g. Bilous et al. 2016), rapid profile evolution (e.g. Pilia et al. 2016), and changing polarisation fractions (e.g. Johnston et al. 2008; Noutsos et al. 2015). 

Pulsar magnetospheres are also both stable and dynamic. Average pulse profile shapes are remarkably stable when summed over at least a few hundred rotations, yet some pulsars' emission has been observed to vary over almost all accessible timescales, manifest in phenomena such as subpulse drifting, nulling, and mode-changing. These various emission phenomena may be connected and allow us to explore the behaviour of the pulsar magnetosphere.

PSR B0823+26, discovered in 1968, exhibits a plethora of emission phenomena over a range of timescales, including subpulse drifting and nulling (e.g. Weltevrede et al. 2007; Young et al. 2012). Sobey et al. (2015) observed PSR B0823+26 using LOFAR on three occasions over several hours, also using three other radio telescopes simultaneously on the third epoch: Westerbork, Lovell, and Effelsberg. During the first LOFAR observation, the pulsar switched from (what we later assumed to be) a `quiet' (Q) emission mode into the more regularly-emitting `bright' (B) mode within one single pulse and with a high circular polarisation fraction.  In the second (using increased bandwidth), we discovered weak and sporadic emission, a Q mode, that is over 100 times weaker than that of the B mode. The Q mode has different properties to the B mode: emission occurs only within the main pulse window, towards slightly later pulse phases; a large nulling fraction 40$\times$ that of the B mode; no detectable subpulse drifting; and a power-law pulse energy distribution (in contrast to a lognormal distribution in the B mode).  The third multi-frequency observations showed that the emission mode switches concurrently across the range of frequencies (a factor of $\sim$20) observed. Therefore, the process appears to be broadband and to reflect a rapid, global change in the structure of the magnetosphere.  Similar behaviour has been observed from other pulsars, such as PSR B0943+10, which switches emission modes in radio and X-rays simultaneously, further alluding to a global magnetospheric change (e.g. Hermsen et al. 2013). 

Within the theoretical framework of force-free magnetospheres, multiple stable solutions with different closed field-line region sizes exist (e.g. Timokhin 2010). The B and Q emission modes with distinctive behaviours may arise from the magnetosphere oscillating between two such states. The mechanism behind the timescales over which these emission phenomena occur (nulling and changing mode within one single pulse, to spending hours--years in one emission state) is not yet well understood.

Only a few dozen pulsars are known to exhibit mode-changing. Long-track (polarisation) observations and timing campaigns permit investigation of longer timescale emission variability in pulsars similar to PSR B0823+26. Discovering and studying pulsars that emit in a Q-mode-like state over longer timescales may require telescopes with greater instantaneous sensitivity, such as the SKA.  Further understanding the magnetospheric emission from pulsars will also assist in more effectively using their signals as probes of, e.g., the Galactic magnetic field and gravitational waves.


 \section{Probing the 3-D Galactic magnetic field structure}

Our current understanding of the large-scale magnetic field structure in the Milky Way disc is that it is broadly aligned with the spiral arms and has a strength of a few $\upmu$G (e.g. Haverkorn et al. 2015). 
Recent analyses often favour an axisymmetric spiral structure with a single reversal in the field direction (e.g. Van Eck et al. 2011), although others advocate multiple reversals (e.g. Han et al. 2006). The 3-D field in the halo is even less well understood, with several proposed geometries (see Haverkorn et al. 2015).

Pulsars are excellent probes of the ionised interstellar medium. Determining dispersion and Faraday rotation measures (DMs and RMs, respectively) from their emission provides an efficient method for ascertaining the average magnetic field strength and direction (parallel to the line of sight) in the intervening media. Pulsars are located throughout the Galaxy, with distances estimated using a Galactic electron density model (e.g. Yao et al. 2017) or measured via parallax. A large set of pulsars allows us to probe the 3-D structure of the Galactic magnetic field (GMF; e.g. Manchester 1972; Manchester 1974; Rand \& Lyne 1994; Han et al. 2006; Noutsos et al. 2008).

Low-frequency observations provide precise measurements of DMs and RMs, due to the wavelength-squared dependence on the pulse arrival delay and polarisation angle rotation, respectively. LOFAR observations produce high-quality polarisation profiles of pulsars at low frequencies ($\le$200\,MHz; Noutsos et al. 2015). Data for a large set of pulsars are available, including the high-band antenna (HBA) `slow' pulsar census (Bilous et al. 2016) and the millisecond pulsar (MSP) census (Kondratiev et al. 2016).  We capitalise on the RM precision potential of these low-frequency data ($\approx$110--190\,MHz) by employing the powerful method of RM-synthesis (Burn 1966; Brentjens \& de Bruyn 2005). The FWHM of the RM spread function in Faraday space ($\delta\phi$; analogous to a point spread function in sky coordinates) is $\approx$1\,rad m$^{\rm{-2}}$ (compared to $\approx$300\,rad m$^{\rm{-2}}$ at 1.3--1.5\,GHz).

The ionosphere is also a magneto-ionic medium and causes additional time and spatial dependence on the observed RM. We currently use the \textsc{ionFR} code (Sotomayor-Beltran et al. 2013) to model the ionospheric RM and correct the observed RM.  This is generally the most substantial contribution to the uncertainties ($\approx$0.1\,rad m$^{\rm{-2}}$) at low frequencies. 

Our Faraday picture of the Galaxy includes 41,632 RMs from extragalactic sources (Oppermann et al. 2015) that provide data for the entire line-of-sight.
The pulsar catalogue (v.\,1.56; Manchester et al. 2005) contains 2613 pulsars, 732 of which have published RMs (28\%). 
To date, we have obtained a catalogue of precise (ionosphere-corrected) low-frequency RMs towards $\approx$200 pulsars using LOFAR (Sobey et al. in prep.). Approximately 90 have no published values, and the remaining are $\approx$30$\times$ more accurate compared to published values, generally from data above 1\,GHz.  

The MWA routinely observes pulsars (e.g. Bhat et al. 2016; McSweeney et al. 2017), the polarisation profiles from which are being verified (Xue et al. in prep.), enabling RM measurements (Sobey et al. in prep.). MWA data also supplies excellent resolution in Faraday space: $\delta\phi\approx$0.4--6.2 rad m$^{\rm -2}$, depending on the centre frequency (89--216\,MHz). RMs towards pulsars have also been determined from polarisation images (e.g. Lenc et al. 2017). The MWA and LOFAR provide complementary resolution in Faraday space and facilitate an all-sky picture of low-frequency RMs.

DM and RM measurements towards pulsars further our knowledge of the 3-D GMF structure. The reconstruction is increasingly hampered by the lack of independent distance measurements, e.g., parallaxes, and the uncertain distance estimates. We are entering an era where precision DMs and RMs are becoming routine, allowing us to monitor temporal and spatial variations and to study, e.g., small-scale turbulent ISM structures, the heliosphere (Howard et al. 2016; Tiburzi et al., these proceedings), and the ionosphere. 


\section{Summary, and the next 50 years }
  
We have described how low-frequency polarisation observations of pulsars assist in the investigations of magnetic fields in astrophysical plasmas, especially pulsar magnetospheres and the ISM, which span approximately 18 orders of magnitude in field strength.  The recent low-frequency renaissance has reinvigorated and expedited this field. Multi-frequency polarisation profiles (e.g. Noutsos et al. 2015) and long-track observations to characterise emission phenomena (e.g. Sobey et al. 2015) contribute towards further understanding the magnetospheric emission mechanism. Low-frequency data facilitate precise DM and RM measurements, demonstrated by a large catalogue of RMs (corrected for ionospheric Faraday rotation) using LOFAR observations, and a growing complementary catalogue using the MWA. These data allow us to investigate the 3-D GMF -- especially the large-scale field towards the halo. 
Ongoing pulsar surveys also assist in this by discovering more pulsars and, therefore, lines-of-sight. 

The SKA's pulsar observing capabilities will revolutionise studies of astrophysical magnetic fields. The strategy of ``find `em, time `em (with full polarisation), and VLBI `em'' (Keane, these proceedings) will facilitate significant advances in our knowledge of magnetospheric emission characteristics and the 3-D GMF structure on large and small scales. 
A good understanding of the antennas' polarisation performance and accurate ionospheric monitoring will be invaluable for these science cases. 
These data will be complementary to the `RM grid' of extragalactic point sources and diffuse Galactic synchrotron images, towards a more comprehensive picture of the GMF (Haverkorn et al. 2015).  

\begin{acknowledgements}


Thank you to all of our collaborators. 
Results discussed are based on observations using the Effelsberg Radio Telescope, the International LOFAR Telescope, the Lovell Telescope, the Murchison Widefield Array, and the Westerbork Radio Synthesis Telescope. 
LOFAR (van Haarlem et al. 2013) is the Low Frequency Array designed and constructed by ASTRON. 
This scientific work makes use of the Murchison Radioastronomy
Observatory, operated by CSIRO. We acknowledge the
Wajarri Yamatji people as the traditional owners of the Observatory
site. 

\end{acknowledgements}

\end{document}